\documentclass[twocolumn,aps,showpacs,preprintnumbers,amsmath,amssymb, superscriptaddress]{revtex4-1}
\pdfoutput=1
\usepackage{bm}
\usepackage{graphicx}
\usepackage{color}
\usepackage{xspace}
\definecolor{dblue}{rgb}{0,0,0.6}
\definecolor{dred}{rgb}{0.9,0,0}
\definecolor{dgreen}{rgb}{0,0.4,0}

\begin{document}

\title{Raman scattering investigation of superconducting Ba$_{2}$Ti$_{2}$Fe$_{2}$As$_{4}$O}

\author{S. -F. Wu}
\affiliation{Beijing National Laboratory for Condensed Matter Physics, and Institute of Physics, Chinese Academy of Sciences, Beijing 100190, China}
\author{P. Richard}\email{p.richard@iphy.ac.cn}
\affiliation{Beijing National Laboratory for Condensed Matter Physics, and Institute of Physics, Chinese Academy of Sciences, Beijing 100190, China}
\affiliation{Collaborative Innovation Center of Quantum Matter, Beijing, China}
\author{W. -L. Zhang}
\affiliation{Beijing National Laboratory for Condensed Matter Physics, and Institute of Physics, Chinese Academy of Sciences, Beijing 100190, China}
\author{C. -S. Lian}
\affiliation{Beijing National Laboratory for Condensed Matter Physics, and Institute of Physics, Chinese Academy of Sciences, Beijing 100190, China}
\author{Y. -L. Sun}
\affiliation{Department of Physics, State Key Lab of Silicon Materials, and Center of Electron Microscopy, Zhejiang University, Hangzhou 310027, China}
\author{G. -H. Cao}
\affiliation{Department of Physics, State Key Lab of Silicon Materials, and Center of Electron Microscopy, Zhejiang University, Hangzhou 310027, China}
\author{J. -T. Wang}
\affiliation{Beijing National Laboratory for Condensed Matter Physics, and Institute of Physics, Chinese Academy of Sciences, Beijing 100190, China}
\author{H. Ding}\email{dingh@iphy.ac.cn}
\affiliation{Beijing National Laboratory for Condensed Matter Physics, and Institute of Physics, Chinese Academy of Sciences, Beijing 100190, China}
\affiliation{Collaborative Innovation Center of Quantum Matter, Beijing, China}

\date{\today}

%\begin{minipage}[t]{6.8in}
\begin{abstract}
We have performed polarized Raman scattering measurements on the newly discovered superconductor Ba$_{2}$Ti$_{2}$Fe$_{2}$As$_{4}$O$_{2}$ ($T_c = 21$ K). We observe seven out of eight Raman active  modes, with frequencies in good accordance with first-principle calculations. The phonon spectra suggest neither strong electron-phonon nor spin-phonon coupling and vary only slightly with temperature, except for one E$_g$ mode associated with large displacements of As atoms near the Ti$_2$O planes. We also identify a small anomaly around 125 K in the linewidth of a A$_{1g}$ mode involving the same As atoms. Our results suggest that the transition at 125 K is most likely driven by electronic interactions taking place in the Ti$_2$O planes.
\end{abstract}


%74.25.nd 	Raman and optical spectroscopy
%74.70.Xa 	Pnictides and chalcogenides
%74.25.Kc 	Phonons

%\end{minipage}
\maketitle

Despite a low superconducting critical temperature ($T_c$) of 1.2 K \cite{Yajima_JPSJ81}, BaTi$_2$Sb$_2$O is widely regarded as a possible candidate for unconventional superconductivity. Indeed, its square Ti$_2$O layers with a Ti 3$d^1$ electron shell are quite similar to the CuO$_2$ layers with a Cu 3$d^9$ electron shell of the high-$T_c$ cuprate superconductors. Interestingly, this material exhibits an anomaly in the electrical resistivity at a temperature $T_a=50$ K, which is attributed either to a charge-density-wave (CDW) or a spin-density-wave (SDW). The observation of similar anomalies at 200 K and 114 K in non-superconducting BaTi$_2$As$_2$O \cite{XF_Wang_JPCM22} and Na$_2$Ti$_2$Sb$_2$O \cite{Adam_ZAAC584}, and its disappearance in Na-doped Ba$_{1-x}$Na$_{x}$Ti$_{2}$Sb$_{2}$O with an enhanced $T_c$ of 5.5 K \cite{Doan_JACS134}, suggest a competition between superconductivity and the CDW/SDW order. Interest for this topic raised even further with the recent discovery of Ba$_2$Ti$_2$Fe$_2$As$_4$O with $T_{c}=21$ K \cite{YL_Sun_JACS134}, which is an inter-growth of BaTi$_2$As$_2$O and BaFe$_2$As$_2$, the parent compound of the 122 family of ferropnictide superconductors. For this material, a transition at 125 K has been identified from the electrical resistivity and the magnetic susceptibility. Unfortunately, whether this transition is related to a CDW or a SDW is still under debate.

In this paper, we use Raman scattering to investigate directly the crystallographic structure of Ba$_{2}$Ti$_{2}$Fe$_{2}$As$_{4}$O. We observed seven out of eight Raman active modes, with frequencies in good accordance with first-principle calculations. The phonon spectra suggest neither strong electron-phonon nor spin-phonon coupling and vary only slightly with temperature, except for a 11 cm$^{-1}$ shift observed for one E$_g$ mode involving large displacements of As atoms near the Ti$_2$O. Our analysis of the phonon linewidths also revealed a small anomaly around $T_a=125$ K for a A$_{1g}$ mode involving the same As atoms. Our results suggest that the transition at $T_a$ is of electronic nature and originates from the Ti$_2$O planes. 

%II.Materials and methods
The Ba$_{2}$Ti$_{2}$Fe$_{2}$As$_{4}$O single crystals used in our Raman scattering measurements were grown by solid-state reactions \cite{YL_Sun_JACS134}. The resistivity of the samples was measured with a Quantum Design physical properties measurement system (PPMS), and the magnetization of the samples was measured with a vibrating sample magnetometer (VSM). The magnetization measurements show that only about 30-40\% of the volume was superconducting, suggesting inhomogeneous superconductivity. The crystals were cleaved in air to obtain flat surfaces and then transferred into a low-temperature cryostat ST500 (Janis) for the Raman measurements between 5 and 300 K with a working vacuum better than $8\times 10^{-7}$ mbar. Raman scattering measurements were performed using 488.0 nm, 514.5 nm and 647.1 nm laser lines in a back-scattering micro-Raman configuration with a triple-grating spectrometer (Horiba Jobin Yvon T64000) equipped with a nitrogen-cooled CCD camera. In this manuscript, we define $x$ and $y$ as the directions along the $a$ and $b$ axes, oriented at 45$^{\circ}$ from the Fe-Fe bounds, and $x'$ and $y'$ as the directions along the Fe-Fe bounds. The $z$ direction corresponds to the $c$-axis perpendicular to the FeAs and TiO$_2$ planes.

\begin{figure}[!t]
\begin{center}
\includegraphics[width=3.4in]{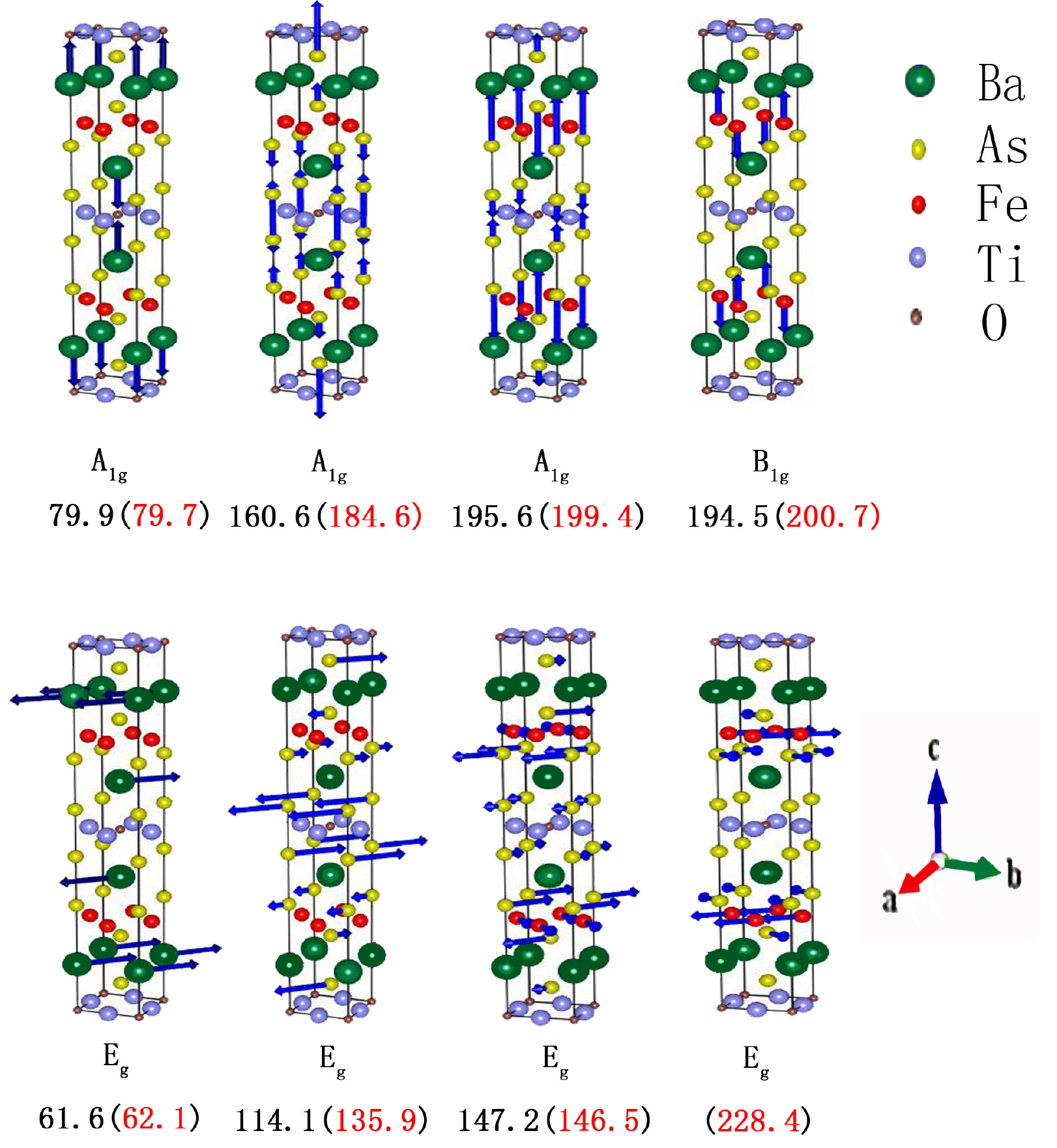}
\end{center}
\caption{\label{vibrate}(Color online). Atomic displacement patterns for the eight Raman-active modes of Ba$_2$Ti$_2$Fe$_2$As$_4$O. The values in black correspond to the experimental values while the ones in red, inside brackets, are the calculated values.}
\end{figure}

The Ba$_2$Ti$_2$Fe$_2$As$_4$O crystal structure is characterized by the space group D$_{4h}^{17}$ (I4/mmm). A simple group symmetry analysis \cite{Comarou_Bilbao} indicates that the phonon modes at the Brillouin zone (BZ) center decompose into [3A$_{1g}$+B$_{1g}$+4E$_{g}$]+[5A$_{2u}$+B$_{2u}$+6E$_{u}$]+[A$_{2u}$+E$_{u}$], where the first, second and third terms represent the Raman-active modes, the infrared(IR)-active modes and the acoustic modes, respectively. To get estimates on the phonon frequencies, we performed first-principle calculations in the non-magnetic phase of the phonon modes at the BZ center in the framework of the density functional perturbation theory (DFPT) \cite{Baroni_RMP73}, using the experimental lattice parameters $a=b=4.0276$ \AA\xspace and $c=27.344$ \AA, and the Wyckoff positions Ba 2i, Ti 4e, Fe 2i, As 2i, O 4e. For all calculations, we used the Vienna \emph{ab initio} simulation package (VASP) \cite{KressePRB54} with the generalized gradient approximation (GGA) of Perdew-Burke-Ernzerhof for the exchange-correlation functions \cite{Perdew_PRB46}. The projector augmented wave (PAW) \cite{Blochl_PRB50} method was employed to describe the electron-ion interactions. A plane wave cutoff energy of 520 eV was used with a uniform $9\times 9\times 9$ Monkhorst-Pack $k$-point mesh for integrations over the BZ. The frequencies and displacement patterns of the phonon modes were derived from the dynamical matrix generated by the DFPT method. The calculated frequencies and the optical activity of the phonon modes are given in Table \ref{EXP_CAL_comparsion}, and the atomic displacements of the Raman-active phonon modes are illustrated in Fig. \ref{vibrate}.

%IV Raman Phonon Results

\begin{figure}[!t]
\begin{center}
\includegraphics[width=3.4in]{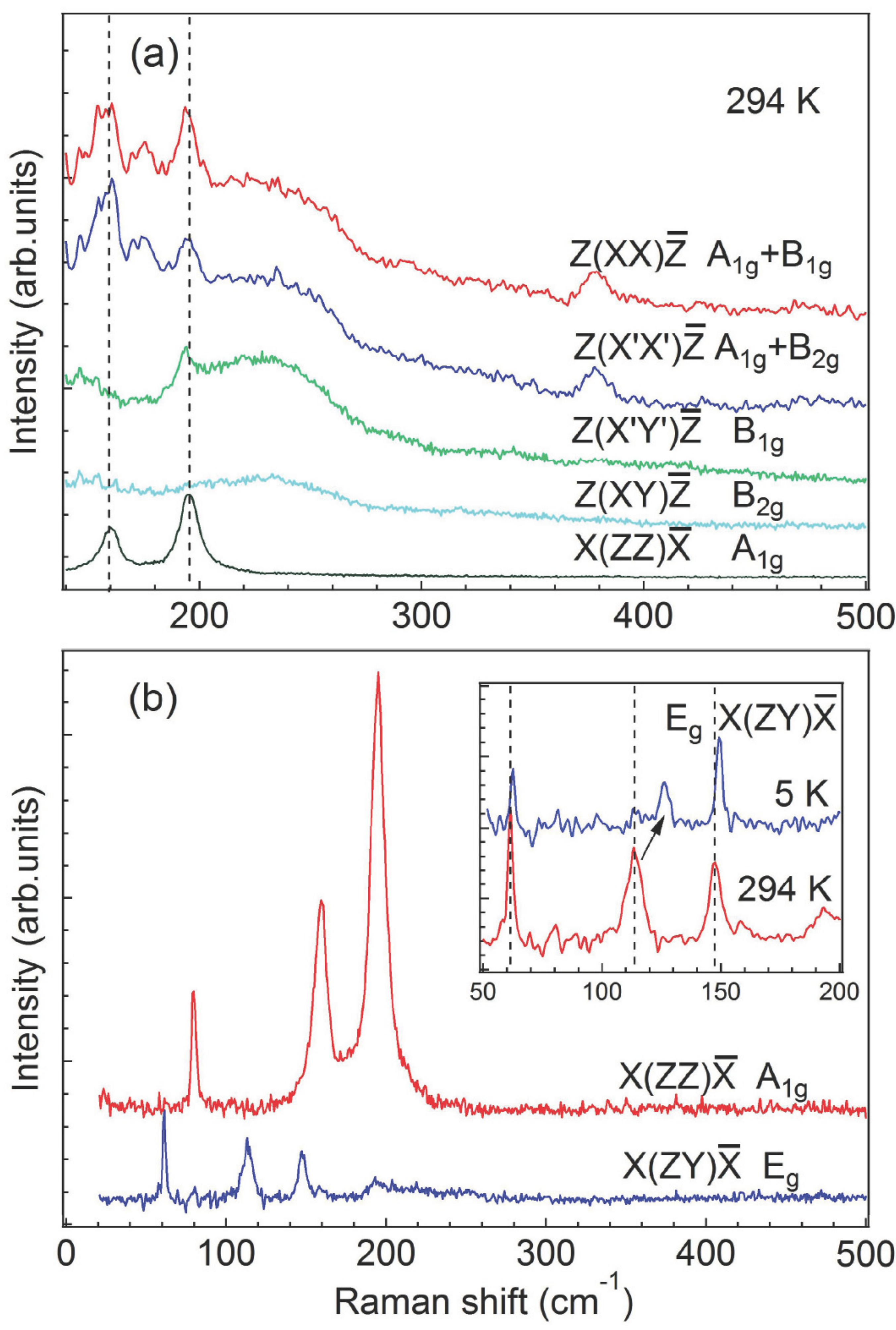}
\end{center}
\caption{\label{phonons_fig}(Color online). (a) Polarization dependence of the $ab$ plane measurements of Ba$_2$Ti$_2$Fe$_2$As$_4$O at 294 K. (b) Polarization dependence of the $ac$ plane measurements at 294 K. The inset compares the Raman spectra recorded at 294 K in the E$_g$ channel. The vertical dashed lines in (a) and in the inset of panel (b) are guides to the eye while the small arrow in the inset indicates a peak shift.}
\end{figure}

In Figs. \ref{phonons_fig}(a) and \ref{phonons_fig}(b), we show the Raman spectra of Ba$_2$Ti$_2$Fe$_2$As$_4$O recorded at room temperature under various polarization configurations. The symmetry of the modes observed are determined by the Raman tensors corresponding to the $D_{4h}$ symmetry group, which are expressed as:

\begin{displaymath}
\textrm{A$_{1g}$=}
\left(\begin{array}{ccc}
a & 0 &0\\
0 & a &0\\
0 & 0 &b
\end{array}\right)
, \textrm{B$_{1g}$=}
\left(\begin{array}{ccc}
c & 0 &0\\
0 & -c &0\\
0 & 0 &0\\
\end{array}\right),
\end{displaymath}
\begin{displaymath}
\textrm{B$_{2g}$=}
\left(\begin{array}{ccc}
0 & d &0\\
d & 0 &0\\
0 & 0 &0\\
\end{array}\right),
\textrm{E$_{g}$=}
\left\{\left(\begin{array}{ccc}
0 & 0 &0\\
0 & 0 &e\\
0 & e &0\\
\end{array}\right),
\left(\begin{array}{ccc}
0 & 0 &-e\\
0 & 0 &0\\
-e & 0 &0\\
\end{array}\right)
\right\}.
\end{displaymath}

\noindent For perfectly aligned crystals, pure A$_{1g}$ symmetry is obtained in the $x(zz)\bar{x}$ configuration. In this channel, we detect 3 sharp peaks at 79.7 cm$^{-1}$, 160.6 cm$^{-1}$ and 195.6 cm$^{-1}$. As illustrated in Fig. \ref{vibrate}, they correspond to $c$-axis vibrations of the Ba atoms, As atoms in near the Ti$_2$O layer (As$^1$) and As atoms in the Fe-As layer (As$^2$), respectively. In Fig. \ref{phonons_fig}(a), we show that the A$_{1g}$ peaks survive in the $z(xx)\bar{z}$ and $z(x'x')\bar{z}$ configurations, for which the A$_{1g}$ signal is mixed with signals from the B$_{1g}$ and B$_{2g}$ channels, respectively.

In agreement with our group analysis, no phonon peak is detected in the $z(xy)\bar{z}$ configuration, which corresponds to pure B$_{2g}$ symmetry. However, we observe a broad hump at 230 cm$^{-1}$ that appears in the four in-plane polarization configurations but disappears in the $ac$ plane measurements, and which we relate to impurities or inhomogeneity. In addition to that peak, a phonon peak is detected at 194.5 cm$^{-1}$ in the $z(x'y')\bar{z}$ configuration corresponding to pure B$_{1g}$ symmetry. Although that peak is located almost at the same position as one of the A$_{1g}$ peaks, several arguments suggest that these two features might have different origins. In principle, A$_{1g}$ peaks should be detected only in configurations for which the polarization vectors of the incident and scattered lights are parallel, which is not the case for both the B$_{1g}$ and B$_{2g}$ configurations. In addition, if the presence of a A$_{1g}$ peak in the B$_{1g}$ spectrum was due to misalignment, the same peak should also appear in the B$_{2g}$ spectrum, in contrast to our results. Moreover, according to our calculations, the energy of the B$_{1g}$ mode corresponding to vibrations of the Fe atoms along the $c$ axis is similar to that of one A$_{1g}$ mode.

Finally, three out of four E$_g$ modes are detected in the $x(zy)\bar{x}$ polarization configuration spectrum shown in Fig. \ref{phonons_fig}(b). As illustrated in Fig. \ref{vibrate}, the E$_g$ modes found experimentally at 61.6 cm$^{-1}$, 114.1 cm$^{-1}$ and 147.2 cm$^{-1}$ are associated with in-plane motions of Ba atoms, As$^1$ atoms near the Ti$_2$O layer and As$^2$ atoms in the Fe-As layer, respectively. Besides the regular Raman-active phonon peaks observed (7 out of 8 modes) and the hump at 230 cm$^{-1}$ mentioned above, we observe small peaks at 175 cm$^{-1}$ and at 380 cm$^{-1}$ in both the $z(xx)\bar{z}$ and $z(x'x')\bar{z}$ configurations, which we cannot assign. Indeed, these peaks do not follow the selection rules corresponding to Ba$_2$Ti$_2$Fe$_2$As$_4$O and we thus suspect that they come from an impurity phase, our samples containing 7.5\% of them \cite{YL_Sun_JACS134}.

\begin{table}
\caption{\label{EXP_CAL_comparsion}Comparison of the calculated and experimental phonon modes at 294 K.}
\begin{ruledtabular}
\begin{tabular}{ccccc}
 Sym. &	Act. &	  Exp. &	Cal.	& Main atom displacements\\
\hline
E$_u$   &   IR &             &        51.7& Ti(xy), As$^1$(xy), Fe(-xy), As$^2$(xy)\\
E$_{g}$& 	Raman& 	    61.6 &	     62.1&	Ba(xy)\\
A$_{2u}$& 	IR& 		     &       69.3&	O(z), As$^1$(z)\\
A$_{1g}$& 	Raman& 	    79.9 &	     79.7&	Ba(z)\\
E$_{u}$& 	IR& 		     &       84.9& 	Ba(xy)\\
A$_{2u}$& 	IR& 		     &       87.9&	Ba(z), Fe(z), As$^2$(z)\\
E$_{g}$&    Raman& 	    114.1& 	    135.9& 	As$^1$(xy)\\
E$_{g}$& 	Raman& 	    147.2& 	    146.5& 	As$^2$(xy)\\
E$_{u}$& 	IR& 		     &      165.8&	Ti(xy), As$^1$(-xy)\\
B$_{2u}$& 	IR& 		     &      178.5&	Ti(z)\\
A$_{1g}$& 	Raman& 	    160.6&      184.6&	As$^1$(z)\\
A$_{2u}$& 	IR& 		     &      190.0&  As$^2$(z), O(-z)\\
A$_{1g}$& 	Raman& 	    195.6&      199.4&	As$^2$(z)\\
B$_{1g}$& 	Raman& 	    194.5    &      200.7& 	Fe(z)\\
E$_{u}$& 	IR& 	         &      226.4&	Ti(xy), O(-xy)\\
E$_{g}$& 	Raman& 	    -    &	    228.4&	Fe(xy), As$^2$(xy)\\
A$_{2u}$& 	IR& 		     &      246.4&	Ti(z), As$^1$(-z)\\
A$_{2u}$& 	IR& 		     &      259.5&	Fe(z), As(-z)\\
E$_{u}$& 	IR& 		     &      274.5&	Fe(xy), As(xy)\\
E$_{u}$& 	IR& 		     &      546.1&	Ti(xy), O(-xy)\\

\end{tabular}
\end{ruledtabular}
\begin{raggedright}
As$^1$ next to the Ti$_2$O layer; As$^2$  Fe-As layer\\
\end{raggedright}
\end{table}

\begin{figure}[!t]
\begin{center}
\includegraphics[width=3.4in]{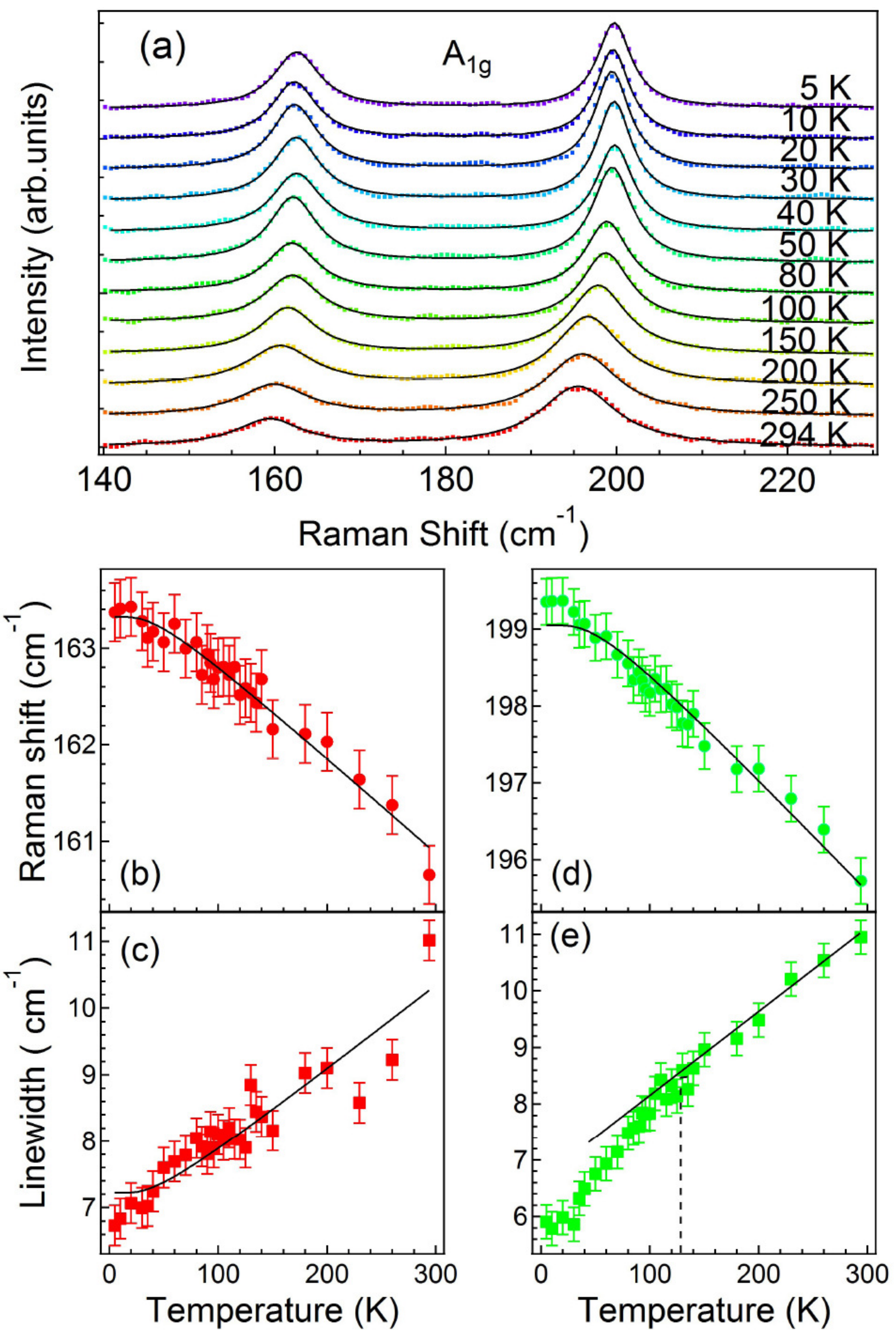}
\end{center}
\caption{\label{T_dependent}(Color online). (a) Temperature dependence of the Raman spectra for the A$_{1g}$ modes at 160.6 cm$^{-1}$ and 195.6 cm$^{-1}$. The black lines represent the fit curves. (b)-(c) Peak position and linewidth (full width at half maximum), respectively, of the A$_{1g}$ mode at 160.6 cm$^{-1}$. (d)-(e) Same as (b)-(c) but for the the A$_{1g}$ mode at 160.6 cm$^{-1}$. The curves in (b)-(d) are fit to the equations Eqs. \eqref{eq_omega} and \eqref{eq_gamma}. The dashed lines in (e) are guides to the eye for an anomaly around 125 K.}
\end{figure}

To investigate the possible role of the lattice on the density-wave transition at $T_a=125$ K and on the superconducting transition at $T_c=21$ K, we cooled the samples down to 5 K. In Fig. \ref{T_dependent}(a), we display the temperature dependence of the A$_{1g}$ peaks at 160.6 cm$^{-1}$ and 195.6 cm$^{-1}$. As expected, the peaks become a little sharper with temperature decreasing. The symmetric Lorentzian lineshapes at all temperatures suggest that there is neither strong electron-phonon coupling nor spin-phonon coupling in this system. In contrast with our expectation for a CDW transition, the peak positions barely change with temperature. We show in Figs. \ref{T_dependent}(b) and \ref{T_dependent}(c) a quantitative analysis of the peak position and the linewidth of the A$_{1g}$ Raman peak at 159.0 cm$^{-1}$, which has been fit simultaneously with two Lorentzian functions convoluted by a Gaussian function representing the system resolution. The peak position $\omega_{ph}(T)$ and the linewidth $\Gamma_{ph}(T)$ follow simple expressions corresponding to the anharmonic phonon decay into acoustic phonons with the same frequencies and opposite momenta \cite{Klemens_PhysRev148,Menendez_PRB29}:

\begin{equation}
\label{eq_omega}
\omega_{ph}(T)=\omega_{0}-C\left( 1+\frac{2}{e^{\frac{\hbar\omega_0 }{ 2k_{B}T}} -1} \right )
\end{equation}
\begin{equation}
\label{eq_gamma}
\Gamma_{ph}(T)=\Gamma_{0}+\Gamma\left( 1+\frac{2}{e^{\frac{\hbar\omega_0 }{ 2k_{B}T}} -1} \right ),
\end{equation}

\noindent where $C$ and $\Gamma$ are positive constants, $\omega_0$ is the bare phonon frequency, and $\Gamma_0$ is a residual, temperature-independent linewidth. From the fit, we extract $\omega_0=163.1$ cm$^{-1}$, $C=0.787$ cm$^{-1}$, $\Gamma=5.00$ cm$^{-1}$ and $\Gamma_0=1.89$ cm$^{-1}$. Similarly, The position of the A$_{1g}$ peak at 195.6 cm$^{-1}$ is well fitted with the formulas with $\omega_0=200.5$ cm$^{-1}$, $C=1.353$ cm$^{-1}$, $\Gamma=4.07$ cm$^{-1}$ and $\Gamma_0=2.20$ cm$^{-1}$. However, it is not possible to fit the linewidth of this peak with Eq. \eqref{eq_gamma}, an increase in the slope while cooling down occurring around $T_a=125$ K, as shown in Fig. \ref{T_dependent}(e).

Our results indicate that neither the phonon peak positions nor the phonon linewidths are affected by the superconducting transition at $T_c=21$ K. Similar behaviors have been reported in Raman studies on NdFe$_2$As$_2$O$_{1-x}$F$_x$ \cite{GallaisPRB78,ZhangL_PRB79}, Sr$_{1-x}$K$_x$Fe$_2$As$_2$ and Ba$_{1-x}$K$_x$Fe$_2$As$_2$ \cite{LitvinchukPRB78,Rahlenbeck_PRB80,KYChoi_JPCM22}, and on Ba(Fe$_{1-x}$Co$_{x}$)$_2$As$_2$ \cite{Chauvilere_PRB80}. On the other hand, anomalies in the peak positions and in the phonon linewidths across the SDW transition have been evidenced in previous Raman studies of BaFe$_2$As$_2$ \cite{Rahlenbeck_PRB80}, CaFe$_2$As$_2$ \cite{Choi_PRB78} and FeTe \cite{Gnezdilov_PRB83}, as well as the splitting of a doubly-degenerate E$_g$ mode \cite{Chauvilere_PRB80} due to the breakdown of the four-fold symmetry accompanying this transition. In contrast, we do not observe any anomaly in the temperature evolution of the peak position of the phonons of Ba$_{2}$Ti$_{2}$Fe$_{2}$As$_{4}$O$_{2}$, and only a small anomaly is observed around $T_a=125$ K in the linewidth of the phonon at 195.6 cm$^{-1}$, as shown in Fig. \ref{T_dependent}(e). This suggests that the transition at $T_a$ is only weakly coupled to the lattice and is most likely from pure electronic origin, which is supported by a spectral weight transfer from the Drude peak to about 800 cm$^{-1}$ in the optical conductivity of the same material across the transition \cite{NL_Wang_private}. This is also consistent with muon spin relaxation measurements on BaTi$_2$(As$_{1-x}$Sb$_x$)$_2$O suggesting the absence of atomic modulation across $T_a$ \cite{Nozaki_PRB88}.

In contrast to the SDW transition found in several Fe-based superconductors, which breaks the four-fold symmetry, we observe no indication of such a symmetry lowering around $T_a$ in Ba$_{2}$Ti$_{2}$Fe$_{2}$As$_{4}$O$_{2}$ that could produce a splitting of the E$_g$ modes, as shown in the inset of Fig. \ref{phonons_fig}(b). Actually, the position of the modes at 61.6 cm$^{-1}$ and 147.2 cm$^{-1}$ barely changes with temperature. Interestingly, only the mode at 114.1 cm$^{-1}$ exhibits a large energy shift as compare to all the other phonon peaks, and is detected at 125 cm$^{-1}$ at 5 K. This is possibly due to a strong coupling between this mode and the transition at $T_a$. As illustrated in Fig. \ref{vibrate}, this E$_g$ mode involves large displacements of the As$^1$ atoms in close proximity to the Ti atoms. Consistently, the A$_{1g}$ mode at 195.6 cm$^{-1}$ showing a linewidth anomaly around $T_a$ also involves the As$^1$ atoms. This suggests that the transition at $T_a$ most likely occurs due to electronic interactions in the Ti$_2$O planes, and that its origin would be closely related to the one found in Fe-free BaTi$_2$As$_2$O \cite{XF_Wang_JPCM22} and Na$_2$Ti$_2$Sb$_2$O \cite{Adam_ZAAC584}. Although more experimental work using probes directly sensitive to the element selective electronic structure is necessary to validate this hypothesis, our work already provides a clear indication that the transition has a purely electronic origin.

In summary, we have performed polarized Raman scattering measurements on the newly discovered superconductor Ba$_{2}$Ti$_{2}$Fe$_{2}$As$_{4}$O$_{2}$ ($T_c = 21$ K). We observe seven out of eight Raman active  modes, with frequencies in good accordance with first-principle calculations. The phonon spectra suggest neither strong electron-phonon nor spin-phonon coupling and vary only slightly with temperature, at the exception of an E$_g$ mode involving large displacements of As atoms in the vicinity of the Ti$_2$O planes. We also report a small anomaly around $T_a$ in the linewidth of an A$_{1g}$ mode also involving As atoms in the vicinity of the Ti$_2$O planes. Our results suggest that the transition at $T_a$ originates from electronic interactions in the Ti$_2$O planes. 

We acknowledge H. Miao and J. -X. Yin for useful discussions. This work was supported by grants from CAS (2010Y1JB6), MOST (2010CB923000,  2011CBA001000, 2011CBA00102, 2012CB821403 and 2013CB921703) and NSFC (11004232, 11034011/A0402, 11234014 and 11274362) from China.

%\bibliography{biblio_short}
\bibliography{biblio_long}

\end{document}